%% file: masking_compumag2009_jrnl_revised.tex
\documentclass[10pt,twocolumn,a4paper]{myIEEEtran}
\usepackage{amsmath}
\usepackage{amsfonts}
\def\R{\mathbb{R}}
\input mycommands.tex

\usepackage{multicol}

  \usepackage[pdftex]{graphicx}
  \DeclareGraphicsExtensions{.pdf,.jpeg,.png}
\usepackage{fixltx2e}
\renewcommand{\baselinestretch}{.98}

\title{\Large\bf
Transformation Optics, Generalized Cloaking and Superlenses%
}
\author{Andr\'e~Nicolet$^1$,~Fr\'ed\'eric~Zolla$^1$,~Christophe~Geuzaine$^2$\medskip \\ \affiliation \\ \Abstract
\thanks{\TheThanks}}

\def\affiliation{
\normalsize
  $^1$ Aix-Marseille Universit\'e, Ecole Centrale de Marseille, CNRS, Institut Fresnel UMR 6133,\\ F-13397 Marseille cedex 20, France\\
  $^2$ University of Li{\`e}ge, Dept. of Electrical Engineering
  and Computer Science, B-4000 Li{\`e}ge, Belgium
}

\def\TheThanks{Corresponding author: A. Nicolet.
E-mail: andre.nicolet@fresnel.fr}

\def\Abstract{\vskip \baselineskip
\quotation
\relax \normalfont \small \bfseries
  In this paper, transformation optics is presented together with a generalization of invisibility cloaking: instead of an empty region of space, an inhomogeneous structure is transformed via Pendry's map in order to give, to any object hidden in the central hole of the cloak, a completely arbitrary appearance. Other illusion devices based on superlenses considered from the point of view of transformation optics are also discussed.

\vskip \baselineskip

\textit{Index Terms}---\,\relax
  Transformation optics, invisibility cloaking, wave propagation, finite element modelling, superlens.
\endquotation
\vspace*{-2\baselineskip}

}

\begin{document}

\maketitle
\def\thepage{\relax}
\markboth{}{}

\markboth{Journal of \LaTeX\ Class Files,~Vol.~6, No.~1, January~2007}%
{Shell \MakeLowercase{\textit{et al.}}: Bare Demo of IEEEtran.cls for Journals}

\section{Introduction}

{In} 2006, it was suggested by Pendry et al. \cite{pendrycloak} that an
object surrounded by a coating consisting of an exotic
material could become invisible to electromagnetic
waves. This device was named ``invisibility cloak'' in reference to Harry Potter, the popular character of J.K. Rowling. Beside his famous cloak, the little wizard has other spells to go unnoticed. Among the most spectacular is the ``polyjuice potion'' that is able to turn somebody into anybody else's appearance \cite{jkrowling}. In this paper, we do not present a potion but rather an optical device able to accomplish the same task, i.e. to give an arbitrary optical response chosen in advance to any other object placed inside the device. In fact, the principle is here very similar to the design of Pendry's invisibility cloak but, instead of geometrically transforming an empty domain, we transform a region containing the object to be imitated, thus leading to a generalization of cloaking.

\section{Transformation Optics}

In recent years, transformation optics has become a very active new field. It has been popularized through the idea of J.B. Pendry that an invisibility cloak can be designed by transforming space and considering the corresponding equivalent material properties \cite{pendrycloak,opl}. Indeed, it is a deep property of Maxwell's equations that they are purely topological (when written in the proper formalism \cite{nicolet}) and that all the metric aspects can be encapsulated in the electromagnetic material properties. A direct consequence is that any continuous transformation of space can be encoded in an equivalent permittivitty and permeability.
Extending this principle beyond continuous transformations allows to design exotic optical devices such as the invisibility cloak.

Exterior calculus is the most natural formalism to write Maxwell's equation \cite{deschamps,burke}
so that they have the following form (in the harmonic case with a pulsation $\omega$ and complex valued fields):
\begin{equation} \left \lbrace
\begin{array}{l}
 d \mathbf{H}=\mathbf{J} - i \omega \mathbf{D} \\%
 d \mathbf{E}=+i \omega \mathbf{B} \\%
 d \mathbf{D}=\rho \\%
 d \mathbf{B}=0
\end{array}
\right.
\label{Maxwell}\end{equation}
where $d$ is the exterior derivative ($d$ plays the role of $\mathbf{curl}$ in the two first equations and of $div$ in the last two ones, see Appendix
), the
1-forms $\mathbf{E}$, $\mathbf{H}$ are the electric and magnetic fields respectively, the
2-forms $\mathbf{D}$, $\mathbf{B}$, and $\mathbf{J}$ are the electric flux density or
displacement, the magnetic flux density or induction, and the
electrical current density respectively, and the $3$-form $\rho$ is
the electric charge density. The only operator involved is the exterior derivative that is completely independent from the metric.

The metric is involved in the Hodge star operator $*$ (see Appendix
) that is necessary to introduce the constitutive laws of materials (including a void, where, $\mathbf{D}= \varepsilon_0 * \mathbf{E}$ and  $\mathbf{B}= \mu_0 * \mathbf{H}$). It can also be argued that it is in fact these very electromagnetic properties of space that determine the metric \cite{symplectic}. This formalism has proved to be very useful in the context of the numerical solution of Maxwell's equations \cite{nicolet,henrotte}. In this case, in has been shown that the topological structure of the equation can be preserved at the discrete level (for instance in Yee's FDTD algorithm or using Whitney discrete forms as finite elements) while the whole process of approximation is concentrated in the design of the discrete Hodge operator \cite{bossavit}.

\subsection{Change of Coordinates in Maxwell's Equations}\label{chcoord}

In the exterior calculus formalism, the only task associated to changing a coordinate system is to determine an explicit expression for the Hodge star operator \cite{nicolet,pcfbook}.
 A very useful point of view is to consider weak formulations where integrals of volume forms ($3$-forms)
 are built with scalar products of forms, i.e., exterior products together with the Hodge operator acting on one of the factors.

For instance, the wave equation for the electric field (in homogeneous media):
\begin{equation}d (\mu^{-1} *d \mathbf{E}) - \omega^2 \varepsilon
*\mathbf{E}=0 ,\label{equ_E_example}\end{equation}
has the following weak formulation: find $\mathbf{E} \in H(\rot, \Omega) $ such that
\begin{equation}\begin{cases}
\int_\Omega \mu^{-1} * d \mathbf{E} \wedge  d \mathbf{\overline{E}'} d\x
- \omega^2 \int_\Omega \varepsilon * \mathbf{E} \wedge  \mathbf{\overline{E}'} d\x
=0 \; \; ,\cr%
\hfill \forall \mathbf{E'} \in H_0(\rot, \Omega)
\label{equ_E_weak}
\end{cases}
\end{equation}
where $\wedge$ is the exterior product (see Appendix
).

We can use the fact that we know how to write this expression components by components in a Cartesian coordinate
system and that we also know how to transform the derivative and the multiple integrals to determine the action of the Hodge operator in other coordinate systems.

Considering a map from the coordinate system $\{u ,v ,w\}$ to the
coordinate system $\{x ,y ,z\}$ given by the functions $x(u, v,
w)$, $y(u, v, w)$, and $z(u, v, w)$, the transformation of the
differentials is given by:
\begin{equation}
\begin{cases}
 dx = \frac{\partial x}{\partial u} du + \frac{\partial x}{\partial v} dv
 + \frac{\partial x}{\partial w} dw \cr%
 dy =\frac{\partial y}{\partial u} du + \frac{\partial y}{\partial v} dv
 + \frac{\partial y}{\partial w} dw \cr%
 dz = \frac{\partial z}{\partial u} du + \frac{\partial z}{\partial v} dv
 + \frac{\partial z}{\partial w} dw.
\end{cases}
\label{transfo_diff}
\end{equation}
Given a $p$-form expressed in the $\{x ,y ,z\}$ coordinate system,
it suffices to replace the $dx, dy, dz$ by the corresponding
1-forms involving  $du, dv, dw$ in the basis exterior monomials to
obtain the expression of the form in the new coordinate system.
Note that the form travels naturally counter to the current with
respect to the map and this is why this transportation of the
forms from $x,y,z$ to $u,v,w$ is called a pull-back.

This
operation can be defined not only between two coordinate systems
but also between two different manifolds even if they do not have
the same dimensions.

Consider two manifolds (or more simply, two open domains of $\R^m$
and $\R^n$ respectively) $N$ and $M$ and a (regular) map $\varphi$
from $N$ to $M$ such that (for simplicity) $\varphi(N)=M$. The
example above shows that it is very easy to express the
differentials of the coordinates on $M$ in terms of the
differentials of the coordinates on $N$ and therefore to find the
image on $N$ of a 1-form on $M$ given by the dual map $\varphi^*$,
from $M$ to $N$, also called, as indicated above, the pull-back.
In fact, any covariant object such as a $p$-form or a metric can
be pulled back by translating the differentials on $M$ into
differentials on $N$. Defined in this way, the operation commutes
of course with the exterior and tensor products but also with the
exterior derivative and the Hodge star (defined with the
pulled-back metric) \cite{nicolet}.

As for contravariant objects such as vector fields, they
travel forward just like the geometrical domains. Given a vector
$\v$ at a point $\p$ on $N$, it suffices to choose a curve
$\gamma$ going through the point and such that the vector is the
tangent vector to the curve at this point, to take the image of
the curve $\varphi(\gamma)$ on $M$ and the vector tangent to this
curve at the point $\varphi(\p)$ as the image of $\v$. Defined in
this way, the map for vectors from $N$ to $M$, denoted by
$\varphi_*(\v)$ or $d\varphi(\v)$, is called the
differential of $\varphi$ or the push-forward
and it can be extended to any contravariant object.

Another fundamental property of the pull-back is its commutativity
with integration in the sense that, for any form $\alpha$ that is
integrable on a subset $\varphi(\Omega)$ of $M$, which is the
image of a subset $\Omega$ of $N$, one has:
\begin{equation}
\int_{\varphi(\Omega)} \alpha = \int_\Omega \varphi^*(\alpha)
\label{integral}.
\end{equation}

All the information for the pull-back is therefore contained in the Jacobian matrix
$\mathbf{J}$ (or maybe we should say matrix field since it depends
on the point in space considered) in terms of which
Eq. (\ref{transfo_diff}) can be written:

\begin{equation}
\left(%
\begin{array}{c}
  dx\\%
  dy\\%
  dz\\%
\end{array}%
\right)%
=\mathbf{J}%
\left(%
\begin{array}{c}
  du\\%
  dv\\%
  dw\\%
\end{array}%
\right)%
\end{equation}
with
$$
\mathbf{J}(u,v,w) = { \frac{\partial(x,y,z)}{\partial(u,v,w)} }
=\left(%
\begin{array}{ccc}
  \frac{\partial x}{\partial u} & \frac{\partial x}{\partial v} &
  \frac{\partial x}{\partial w}\\
  \frac{\partial y}{\partial u} & \frac{\partial y}{\partial v} &
  \frac{\partial y}{\partial w}\\
  \frac{\partial z}{\partial u} & \frac{\partial z}{\partial v} &
  \frac{\partial z}{\partial w}\\
\end{array}%
\right).%
$$

Using matrix notation, the detailed computation of the relation
between the coefficients of a 1-form $\mathbf{\alpha}$ in $\{x, y, z
\}$ and $\{u, v, w \}$ coordinates is performed as follows:

$$
\begin{array}{ll}
\mathbf{\alpha} =\alpha_x dx + \alpha_y dy + \alpha_z dz =%
(\alpha_x \, \alpha_y \, \alpha_z) \left(%
\begin{array}{c}
  dx\\%
  dy\\%
  dz\\%
\end{array}%
\right)
$$
\\
$$
=(\alpha_x \, \alpha_y \, \alpha_z) \; \mathbf{J} \left(%
\begin{array}{c}
  du\\%
  dv\\%
  dw\\%
\end{array}%
\right)
$$
\\
$$
=\alpha_u du + \alpha_v dv + \alpha_w dw =%
(\alpha_u \, \alpha_v \, \alpha_w) \left(%
\begin{array}{c}
  du\\%
  dv\\%
  dw\\%
\end{array}%
\right)\cr%
\end{array}
$$
and the following relation is obtained:
\begin{equation}
(\alpha_x \, \alpha_y \, \alpha_z) \; \mathbf{J} = (\alpha_u \, \alpha_v \, \alpha_w)
\label{pull1}.\end{equation}%
Now the contributions to weak form integrals like Eq. (\ref{equ_E_weak}) may have the
following form: $$\int_\Omega \mathbf{\alpha} \wedge *\mathbf{\alpha}',
$$ where $\mathbf{\alpha}$ and $\mathbf{\alpha}'$ are 1-forms (that can be
obtained as gradients of a scalar field, although it really does
not matter here).
The question is: how to deal with
the Hodge operator? A direct attack would be to pull back the
metric and use the explicit expression of the operator but it is
faster here to take advantage of the simple form of the scalar
product in Cartesian coordinates that reduces to the dot product.
Again using matrix notation   (where $\mathbf{J}^{-T}$ is the inverse of $\mathbf{J}^{T}$):
\begin{equation}
\begin{array}{ll} \mathbf{\alpha} \wedge *\mathbf{\alpha}'
= (\alpha_x \, \alpha_y \, \alpha_z)(\alpha'_x \, \alpha'_y \, \alpha'_z)^T dx \wedge dy \wedge
dz \cr%
= (\alpha_u \, \alpha_v \, \alpha_w)\mathbf{J}^{-1}[(\alpha'_u \, \alpha'_v \,
\alpha'_w)\mathbf{J}^{-1}]^T dx \wedge dy \wedge dz \cr%
=  (\alpha_u \, \alpha_v \, \alpha_w)\mathbf{J}^{-1}\mathbf{J}^{-T}(\alpha'_u \,
\alpha'_v \, \alpha'_w)^T \det(\mathbf{J}) du \wedge dv \wedge dw. \cr%
\end{array}
\end{equation}
The first line is the definition of the scalar product of $1$-forms equated to the scalar product in Cartesian coordinates.

The fact that the transformation of $3$-forms $dx \wedge dy \wedge dz=$ $\det(\mathbf{J}) du \wedge dv \wedge dw$ only involves the
Jacobian, i.e. the determinant of the Jacobian matrix, has been
used here. Hence, the only difference from the case of Cartesian
coordinates is that one of the (column) vectors has to be
multiplied on the left by a symmetric matrix $\mathbf{T}^{-1}$
before performing the dot product, where $\mathbf{T}$ is given by:
\begin{equation}\mathbf{T} = \frac{\mathbf{J}^T \mathbf{J}}{\det(\mathbf{J})}.\end{equation}

It is now interesting to look at how a particular 2-form basis
monomial transforms, for instance,
 $$dx \wedge dy = [\frac{\partial
x}{\partial u} du + \frac{\partial x}{\partial v} dv
 + \frac{\partial x}{\partial w} dw ] \wedge
 [\frac{\partial y}{\partial u} du + \frac{\partial y}{\partial v} dv
 + \frac{\partial y}{\partial w} dw ]$$ $$= \left(\frac{\partial
x}{\partial u}\frac{\partial y}{\partial v}-\frac{\partial
x}{\partial v}\frac{\partial y}{\partial u}\right) du \wedge dv
 +\left(\frac{\partial
x}{\partial v}\frac{\partial y}{\partial w}-\frac{\partial
x}{\partial w}\frac{\partial y}{\partial v}\right) dv \wedge dw
+$$ $$\left(\frac{\partial x}{\partial w}\frac{\partial y}{\partial
u}-\frac{\partial x}{\partial u}\frac{\partial y}{\partial w} \right) dw
\wedge du. $$
The cofactors of $\mathbf{J}$ are now involved in the
transformation. (These are the elements of
$\mathbf{J}^{-T}\det(\mathbf{J})$.)

Given a $2$-form:
\begin{equation}
\begin{array}{ll}
\mathbf{\beta} &=\beta_x dy \wedge dz + \beta_y dz \wedge dx + \beta_z dx \wedge
dy \cr%
           &=\beta_u dv \wedge dw + \beta_v dw \wedge du + \beta_w du \wedge dv%
\end{array}
\end{equation}
the following relation is obtained:
\begin{equation}
(\beta_x \, \beta_y \, \beta_z) \; \mathbf{J}^{-T}\det(\mathbf{J}) = (\beta_u \,
\beta_v \, \beta_w)
\label{pull2}\end{equation}%
and, considering the scalar product $\mathbf{\beta} \wedge *\mathbf{\beta}'$ of two such $2$-forms, it is straightforward to show that the matrix involved in the transformation of this scalar product is here
$\mathbf{T}$ (instead of its inverse in the case of the scalar product of two $1$-forms):
\begin{equation}
\begin{array}{ll} \mathbf{\beta} \wedge *\mathbf{\beta}'
= (\beta_x \, \beta_y \, \beta_z)(\beta'_x \, \beta'_y \, \beta'_z)^T dx \wedge dy \wedge
dz \cr%
= (\beta_u \, \beta_v \, \beta_w)\frac{\mathbf{J}^{T}}{\det(\mathbf{J})}[(\beta'_u \, \beta'_v \,
\beta'_w)\frac{\mathbf{J}^{T}}{\det(\mathbf{J})}]^T \det(\mathbf{J}) du \wedge dv \wedge dw \cr%
=  (\beta_u \, \beta_v \, \beta_w)\mathbf{J}^{T}\mathbf{J}\frac{1}{\det(\mathbf{J})}(\beta'_u \,
\beta'_v \, \beta'_w)^T  du \wedge dv \wedge dw. \cr%
\end{array}
\end{equation}

Everything can now be summarised in the following recipe that
takes into account implicitly the Hodge star: consider a $3$-form
$\gamma$ to be integrated on a domain $\Omega$ in order to
get $\int_\Omega \gamma$ to contribute to a weak form, then:
\begin{itemize}
  \item If the integrand involves only scalars (0-forms or
  3-forms and it does not matter if the 3-forms are
  expressed as the divergence of a vector field) or if it is the exterior product of a 1-form and a
  2-form (and it does not matter if they are respectively a
  gradient and a curl) looking superficially like a scalar product of vectors,
   only $\det(\mathbf{J})$ has to be introduced as a factor.
  \item If the integrand is the scalar product of two 1-forms (and it does
   not matter if one or both 1-forms are expressed as the gradient of a
   scalar field), multiply on the left one of the column vectors of coefficients
    by the matrix $\mathbf{T^{-1}}$.
  \item If the integrand is the scalar product of two 2-forms (and it does
   not matter if one or both 2-forms are expressed as the curl of a
  vector field), multiply on the left one of the column vectors of coefficients
    by the matrix $\mathbf{T}$.
\end{itemize}

The expression obtained for $\varphi^*(\gamma)$ depending on
  variables  $u$, $v$ and $w$ (coordinates $x$, $y$ and $z$ have been
  replaced by the functions $x(u,v,w)$, $y(u,v,w)$ and $z(u,v,w)$
  respectively) is integrated on $\Omega$ to get the
  desired contribution to the volume integral of the weak formulation.

It can also be interesting to consider a compound transformation,
i.e. the transformation of a transformation. Consider three systems
of coordinates $u_i$, $X_i$ and $x_i$ (possibly on different
manifolds) and the maps $\varphi_{Xu}: u_i \rightarrow X_i$ given
by functions $X_i(u_j)$ and  $\varphi_{xX}: X_i \rightarrow x_i$
given by functions $x_i(X_j)$. The composition map $\varphi_{xX}
\circ \varphi_{Xu}= \varphi_{xu}: u_i \rightarrow x_i $ is given
by the functions: $x_i(X_j(u_k))$. If $\mathbf{J}_{xX}$ and
$\mathbf{J}_{Xu}$ are the Jacobian matrices of the maps
$\varphi_{xX}$ and $\varphi_{Xu}$ respectively, the Jacobian
matrix  $\mathbf{J}_{xu}$ of the composition map $\varphi_{xu}$ is
simply the product of the Jacobian
matrices:$$\mathbf{J}_{xu}=\mathbf{J}_{xX}\mathbf{J}_{Xu}.$$ This
rule naturally applies for an arbitrary number of maps.

It is also worth noting that the matrix $\mathbf{J}^T \mathbf{J}$
is nothing but the metric tensor whose coefficients are expressed
in the local coordinates.

\subsection{The Geometric Transformation - Equivalent Material Principle}

A very interesting interpretation of the preceding formulae is that the matrix $\mathbf{T}$
and its inverse can be viewed as tensorial characteristics of
equivalent materials.

By inspection of Eq. (\ref{equ_E_weak}), it appears that $\mu^{-1}$ is present as a factor in the term involving the exterior derivatives, i.e. a scalar product of two $2$-forms and that the $\mathbf{T}$ factor can be introduced by multiplying $\mu$ by $\mathbf{T}^{-1}$ (and therefore turning it in a tensor quantity).
It appears also that $\varepsilon$ is present as a factor in the term involving directly the electric field, i.e. a scalar product of two $1$-forms and that the $\mathbf{T}^{-1}$ factor can be introduced by multiplying $\varepsilon$ by $\mathbf{T}^{-1}$ (and therefore turning it also in a tensor quantity).

Therefore, the only thing to do in the transformed coordinates to compute the integrals of the weak form
is to replace the materials (often homogeneous and isotropic) by
equivalent ones that are inhomogeneous (their characteristics are no
longer piecewise constant but merely depend on $u,v,w$ coordinates)
and anisotropic ones (tensorial nature) whose properties are given
by
\begin{equation}
\underline{\underline{\varepsilon'}} = \varepsilon \mathbf{T}^{-1}
\; ,  \quad
 \hbox{and} \quad
\underline{\underline{\mu'}}= \mu \mathbf{T}^{-1} \; .
\label{epsmuT}
\end{equation}

We note that there is no change in the impedance of the media
since the permittivity and permeability undergo the same
transformation. As for the vector analysis operator and product,
everything works as if we were in Cartesian coordinates.

 In electromagnetism, changing a material can thus
be viewed as changing metric properties and conversely a change of
coordinates can be taken into account by introducing a fictitious
equivalent material. For a general transformation, the equivalent
material is inhomogeneous and anisotropic. It may be interesting in some cases to introduce
non-orthogonal coordinate systems to facilitate the solution of
particular problems, e.g. helicoidal geometries \cite{twistedEPJ,twistedJWAves}.


It is straightforward to generalize the present rules to initially anisotropic material properties. It has also to be noted that they need not to be initially homogeneous. Therefore, the basic principle of transformation optics can be stated in a very general setting: For all our practical purposes, $M$ and $N$ will be here the whole or parts of $\R^3$.
Given a map $\varphi$ from a space $N$ to a space $M$ determining a geometric transformation (i.e. given a Cartesian coordinate system $\mathbf{x}$ on $M$ and an arbitrary coordinate system $\mathbf{x}'$ on $N$, $\varphi: N \rightarrow M$ is described by $\mathbf{x}(\mathbf{x'})$, i.e. $\mathbf{x}$ given as function of $\mathbf{x}'$),
when one  has an electromagnetic system described by the tensor fields $\underline{\underline{\varepsilon}}(\mathbf{x})$ for the dielectric permittivity and $\underline{\underline{\mu}}(\mathbf{x})$ for the magnetic permeability in the space $M$,  if one replaces the initial material properties by equivalent material properties given by the following rule \cite{pcfbook,milton,twistedEPJ,twistedJWAves}:
      \begin{eqnarray}
      \underline{\underline{\varepsilon'}}(\mathbf{x'}) =
      \mathbf{J}^{-1}(\mathbf{x'})\underline{\underline{\varepsilon}}(\mathbf{x}(\mathbf{x'}))\mathbf{J}^{-T}(\mathbf{x'})\det(\mathbf{J}(\mathbf{x'})),  \notag \\
      \underline{\underline{\mu'}}(\mathbf{x'}) =
      \mathbf{J}^{-1}(\mathbf{x'})\underline{\underline{\mu}}(\mathbf{x}(\mathbf{x'}))\mathbf{J}^{-T}(\mathbf{x'})\det(\mathbf{J}(\mathbf{x'})),
      \label{equivalence_rule}
      \end{eqnarray}
 one gets an equivalent problem on $N$. Here, an equivalent problem means that the solution of the new problem on $N$, i.e. electromagnetic quantities described as differential forms, are the pulled back of the solution \cite{nicolet} of the original problem on $M$ and that the same Maxwell's equations (i.e. as if we were in Cartesian coordinates or, more accurately, having the same form as (\ref{Maxwell}) written with the exterior derivative) are still satisfied.

The transformation rules between $M$ and $N$, for components of the fields that are $1$-forms such as $\mathbf{E}$ and $\mathbf{H}$, is given by Eq. (\ref{pull1}) and, for the components of the fields that are $2$-forms such as $\mathbf{B}$, $\mathbf{D}$, and $\mathbf{J}$, is given by Eq. (\ref{pull2}). It appears clearly that the invariant quantities, according to Eq. (\ref{integral}), are the global quantities built as integral of $p$-forms on $p$ dimensional geometrical objects: the line integrals of the electric field and of the magnetic field along a curve, the fluxes across a surface of the electric displacement, the magnetic flux density, the current density, the Poynting vector...


\section{Generalized cloaking}

In this section, we present a generalization of cloaking able to arbitrarily transform the electromagnetic appearance of an object. The basic principle is to obtain the constitutive relations of the cloak by application of a space transformation to a non-empty region.
Invisibility can be considered as a particular case that corresponds to choosing the empty space as the object to be faked.

\begin{figure}[h]
{\centering  \includegraphics[width=0.45\textwidth]{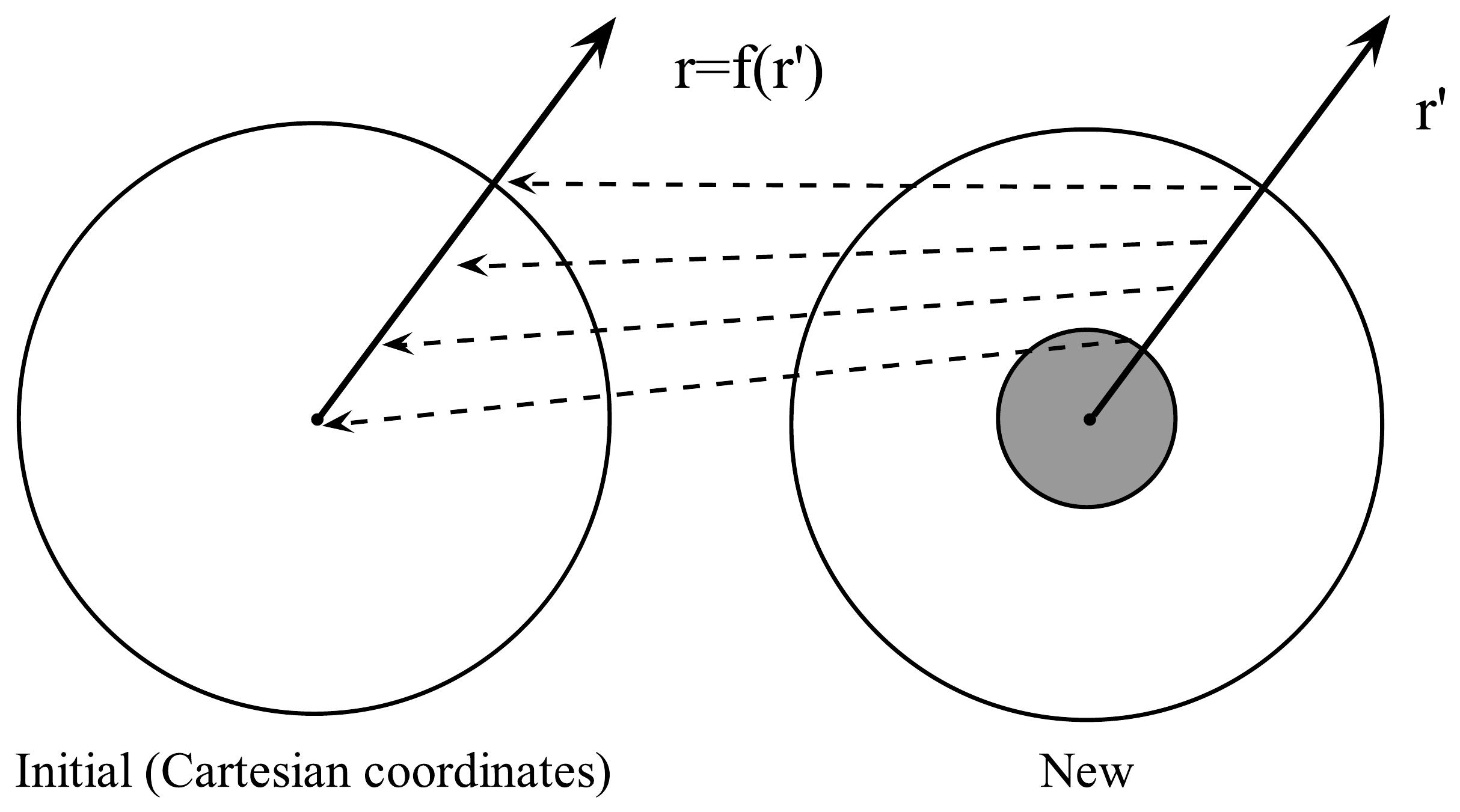}
\caption{\label{tfopen}  Pendry's map of an annulus to a disk used to determine the material properties of an invisibility cloak via the equivalence principles.  }}
\end{figure}

In the case of the cylindrical Pendry's map \cite{pendrycloak,opl,compel_geo}, described by the transformation of the 2D cross section, the plane $\R^2$ minus a disk $D_1$ of radius $R_1$ is mapped on the whole plane $\R^2$ in such a way that a disk $D_2$ of radius $R_2 > R_1$, concentric with $D_1$, is the image of the annulus $D_2\backslash D_1$ by a radial transformation (see Fig. \ref{tfopen}). In cylindrical coordinates, this transformation is given by:

\begin{equation}\begin{cases}  r = (r'-R_1) R_2/(R_2-R_1) \; \mathrm{for} \; R_1 \leq r'\leq R_2, \; \\ \theta = \theta', \; z = z'. \;
\label{pendrysmap}\end{cases}\end{equation}

As for the outside of the disk $D_2$, the map between the two copies of $\R^2\backslash D_2$ is the identity map.

The material properties given by rule (\ref{equivalence_rule}) corresponding to this transformation provide an ideal invisibility cloak: outside $D_2$, everything behaves as if we were in free space, including the propagation of electromagnetic waves across the cloak, and is completely independent of the content of $D_1$.


Now, rule (\ref{equivalence_rule}) may be applied to $D_2$ containing objects with arbitrary electromagnetic properties so that a region cloaked by this device is still completely hidden but has the appearance of the objects originally in $D_2$. We may call this optical effect masking \cite{teixeira} or ``polyjuice'' effect.

\begin{figure}[h]
{\centering  \includegraphics[width=0.45\textwidth]{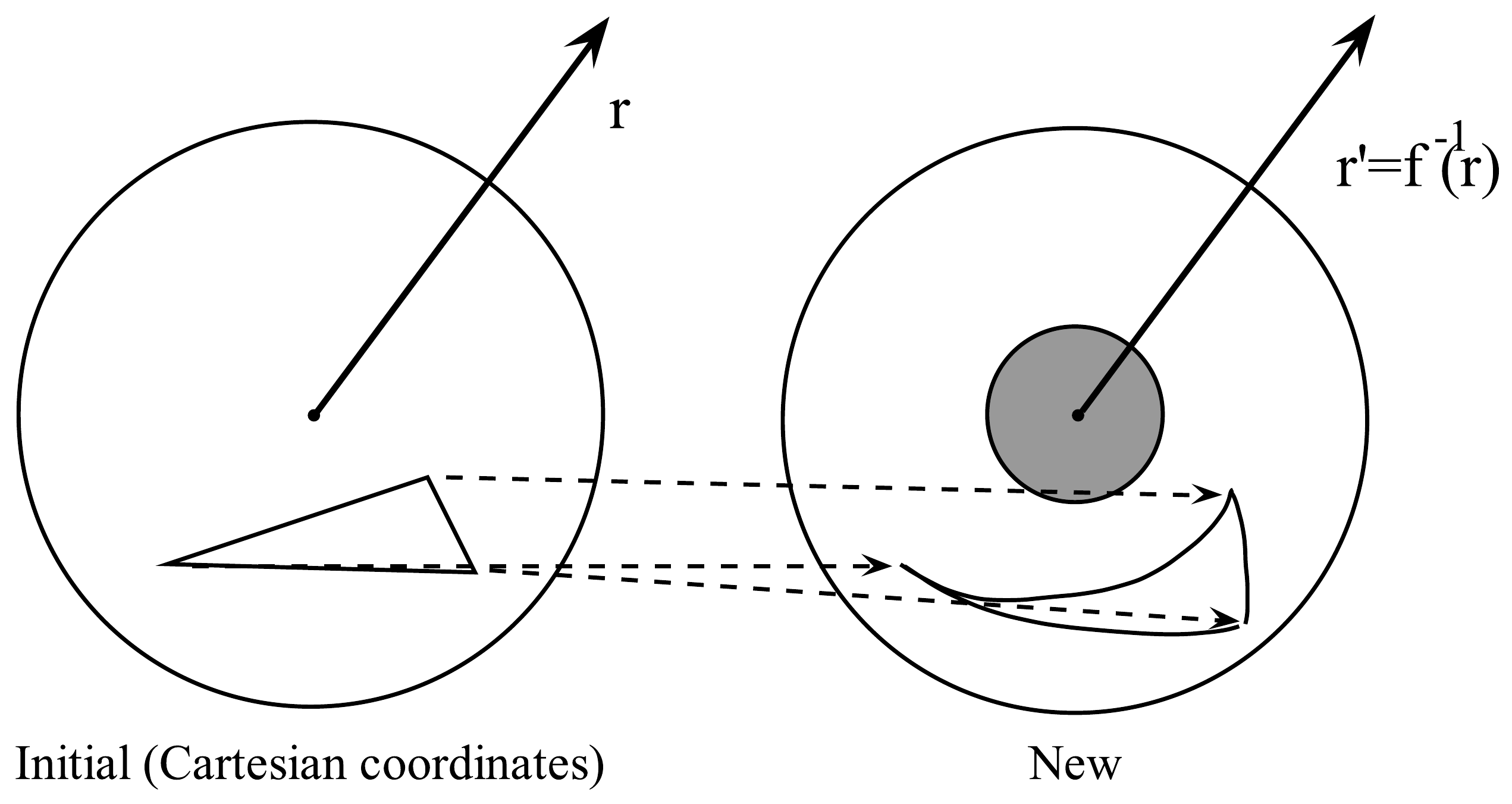}
\caption{\label{tfopush}  When the material properties are piecewise defined, a pushforward of the geometry involving the inverse transformation is useful.  }}
\end{figure}

\section{Numerical modeling}

\begin{figure}[h]
{\centering  \includegraphics[width=0.48\textwidth]{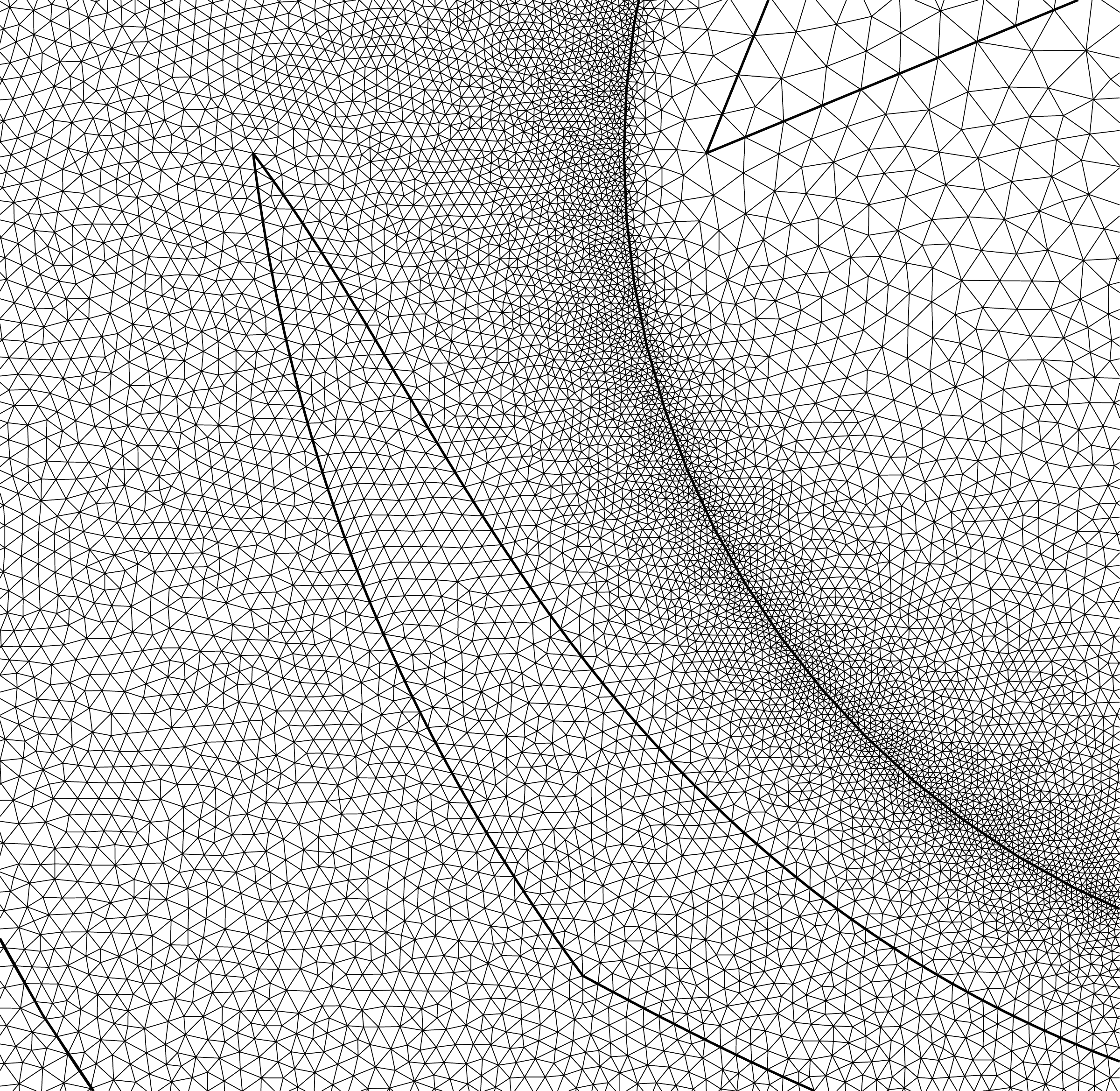}
\caption{\label{mesh}  This figure shows a part of the triangular mesh used for the finite element modeling of the scattering problem of  Fig. \ref{fig2}. The singular behavior of the permittivity and of the permeability requires a very fine mesh along the inner boundary of the cloak in order to achieve a satisfactory accuracy with the numerical model.  }}
\end{figure}

\begin{figure}[h]
{\centering \includegraphics[width=0.48\textwidth]{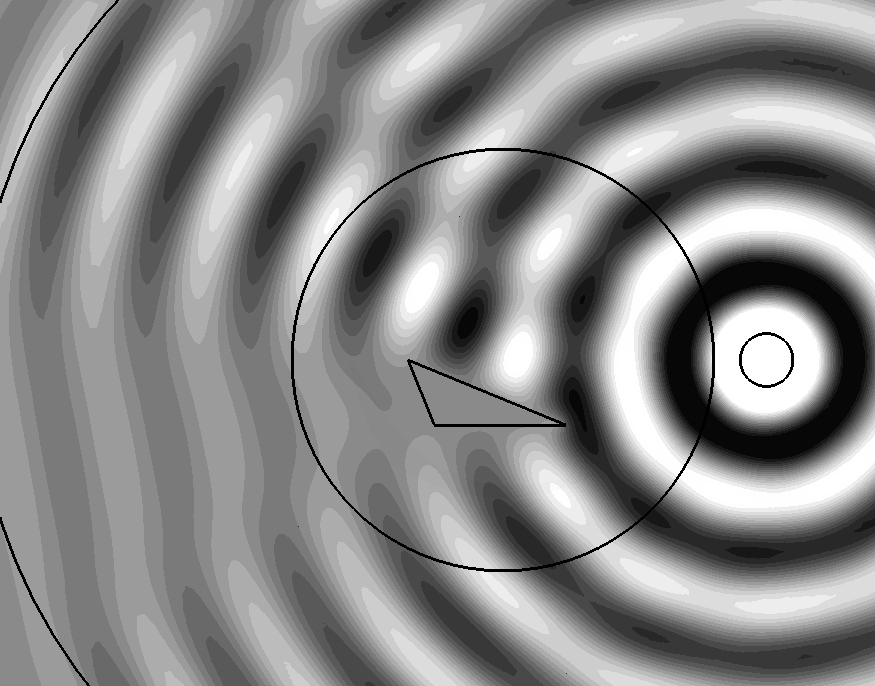}
\caption{\label{fig1}  A conducting triangular cylinder is scattering cylindrical waves.}}
\end{figure}

\begin{figure}[h]
{\centering  \includegraphics[width=0.48\textwidth]{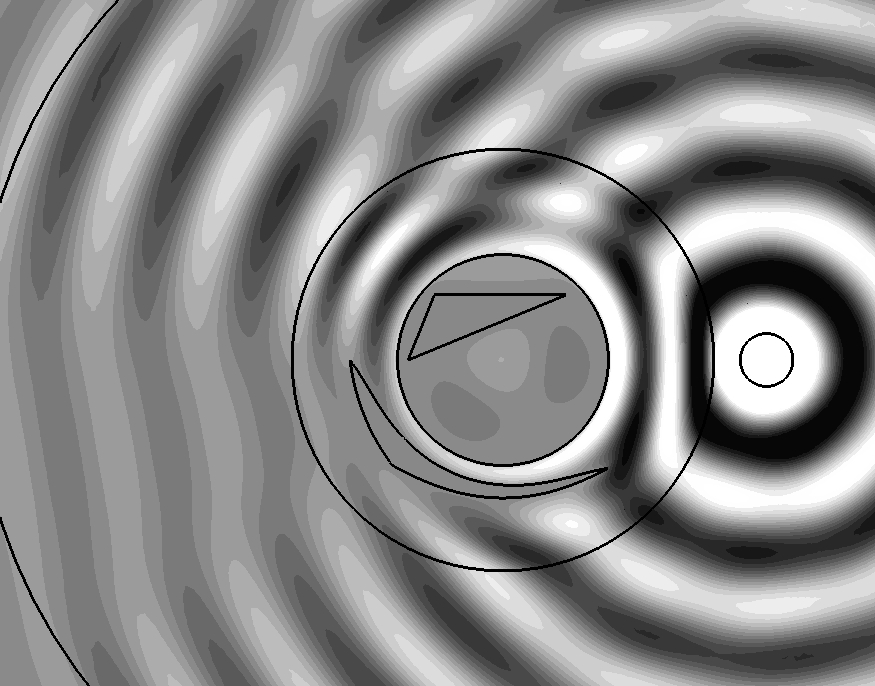}
\caption{\label{fig2}  A triangular cylinder different from the one on Fig. \ref{fig1} is surrounded by a cloak designed to reproduce the scattering pattern of the Fig. \ref{fig1} triangular cylinder in spite of the change of scattering object. Of course, the scattering object inside the cloak may be arbitrary as far as it is small enough to fit inside the cloak.}}
\end{figure}

\begin{figure}[h]
{\centering  \includegraphics[width=0.48\textwidth]{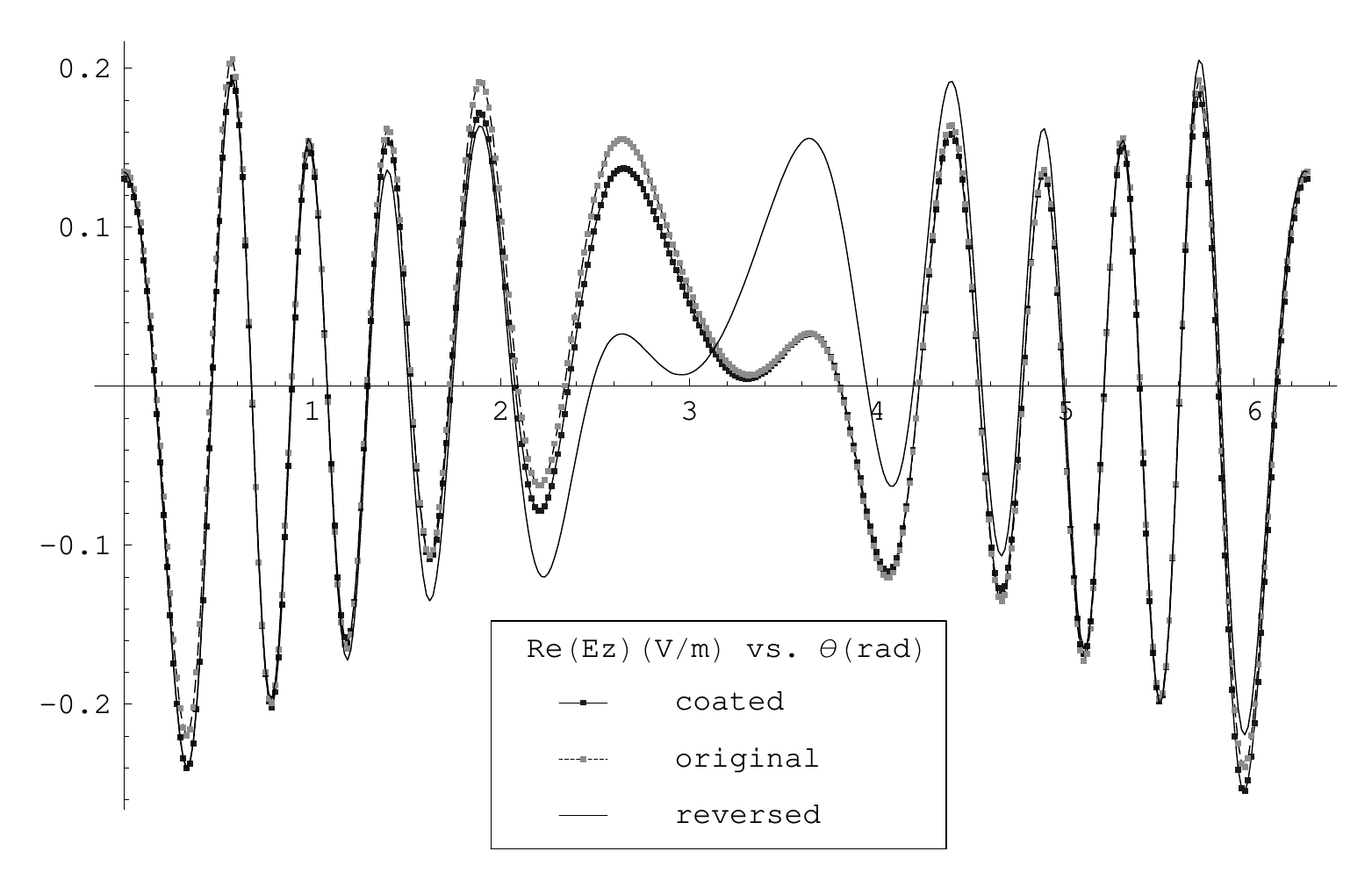}
\caption{\label{fig3}   The value of the electric field (the real part of $E_z$) on a circle of radius $4\lambda$ concentric with the cloak is represented as a function of the position angle $\theta$ (increasing counterclockwise and with $\theta=0$ corresponding to the point the most on the right). The three configurations considered here are the ones of Fig. \ref{fig2} (coated), Fig. \ref{fig1} (original), and the triangle of Fig. \ref{fig2} without the coating (reversed).  } }
\end{figure}


Figs. \ref{fig1} and \ref{fig2} show the effect of masking on a scattering structure. On Fig. \ref{fig1}, a cylindrical TM wave emitted by a circular cylindrical antenna is scattered by a conducting triangular cylinder (the longest side of the cross section is $1.62\lambda$ and $\varepsilon_r=1 + 40 i $). The field map  represents the longitudinal electric field $E_z(x,y)$ and the outer boundary of the cloak is shown to ease the comparison with the masked case. On Fig. \ref{fig2}, the same cylindrical TM wave is scattered by a masked triangular cylinder (but the scattering object inside the cloak may be arbitrary as far as it is small enough to fit inside the cloak). This triangular cylinder is the symmetric of the previous one with respect to the horizontal plane containing the central fibre of the cylindrical antenna. This bare scatterer would therefore give the Fig. \ref{fig1} image inverted upside-down but, here, this object is surrounded by a cloak in order to give the very same scattering as before. Indeed, on both sides, the electric fields outside the cloak limit are alike.

  Fig. \ref{fig3} highlights the different scattering patterns by displaying the value of $\Re e(E_z)$ on a circle of radius $4 \lambda$ located around the antenna-scatterer system in the three following cases: the case of Fig. \ref{fig1} (original) with the triangle alone, the case of Fig. \ref{fig2} (coated), and the triangle of Fig. \ref{fig2} without the coating (reversed). It is obvious that the coating restores the field distribution independently of the object present in the central hole.

The numerical computation is performed using the finite element method (via the free GetDP \cite{getdp} and Gmsh \cite{gmsh} software tools). The mesh is made of 148,000 second order triangles   including the Perfectly Matched Layers used to truncate the computation domain. The singularity of $\varepsilon$ and $\mu$ requires a very fine mesh in the vicinity of the inner boundary of the cloak (see Fig. \ref{mesh}) and is also responsible for the small discrepancies between the numerical model and a perfect cloak (see Fig. \ref{fig2}) --- including the non zero field in the hole of the cloak.

Note that a small technical problem arises in practice when rule (\ref{equivalence_rule}) is applied: the material properties are defined piecewise on various domains and it is very useful to know explicitly the boundaries of these domains, e.g. to build the finite element mesh (see Fig. \ref{mesh}). These boundaries are curves in the cross section and are thus contravariant objects. Therefore, their transformation requires the inverse map (see Fig. \ref{tfopush}) $\varphi^{-1}$ from $M$ to $N$. Fortunately, map (\ref{pendrysmap}) is very simple to invert.
More explicitly, for a given curve $\mathbf{x}(t)$ of parameter $t$ in the initial Cartesian coordinates, its push forward by Pendry's map is:
\begin{equation}
\mathbf{x}'(t)=\varphi^{-1}(\mathbf{x}(t))=(\frac{R_2-R_1}{R_2}+ \frac{R_1}{\|\mathbf{x}(t) \|})\mathbf{x}(t),
\end{equation}
with the same variation of the parameter $t$ and $\|\mathbf{x}\|=r$. Note that the most common curves used in the design of devices, i.e. line segments and arc of circles, are transformed to less usual curves except   for radial segments (with respect to the center of the cloak) and arc of circles concentric with the cloak.

 On Fig. \ref{fig2}, the image by $\varphi^{-1}$ of the triangle of Fig. \ref{fig1} is the curvilinear triangle inside the coating region of the cloak. In practice, this anamorphosis of the triangle is described by three splines interpolating each 40 points that are images of points of the segments by $\varphi^{-1}$.

\section{Superlens Illusion}

\begin{figure}[h]
{\centering  \includegraphics[width=0.3\textwidth]{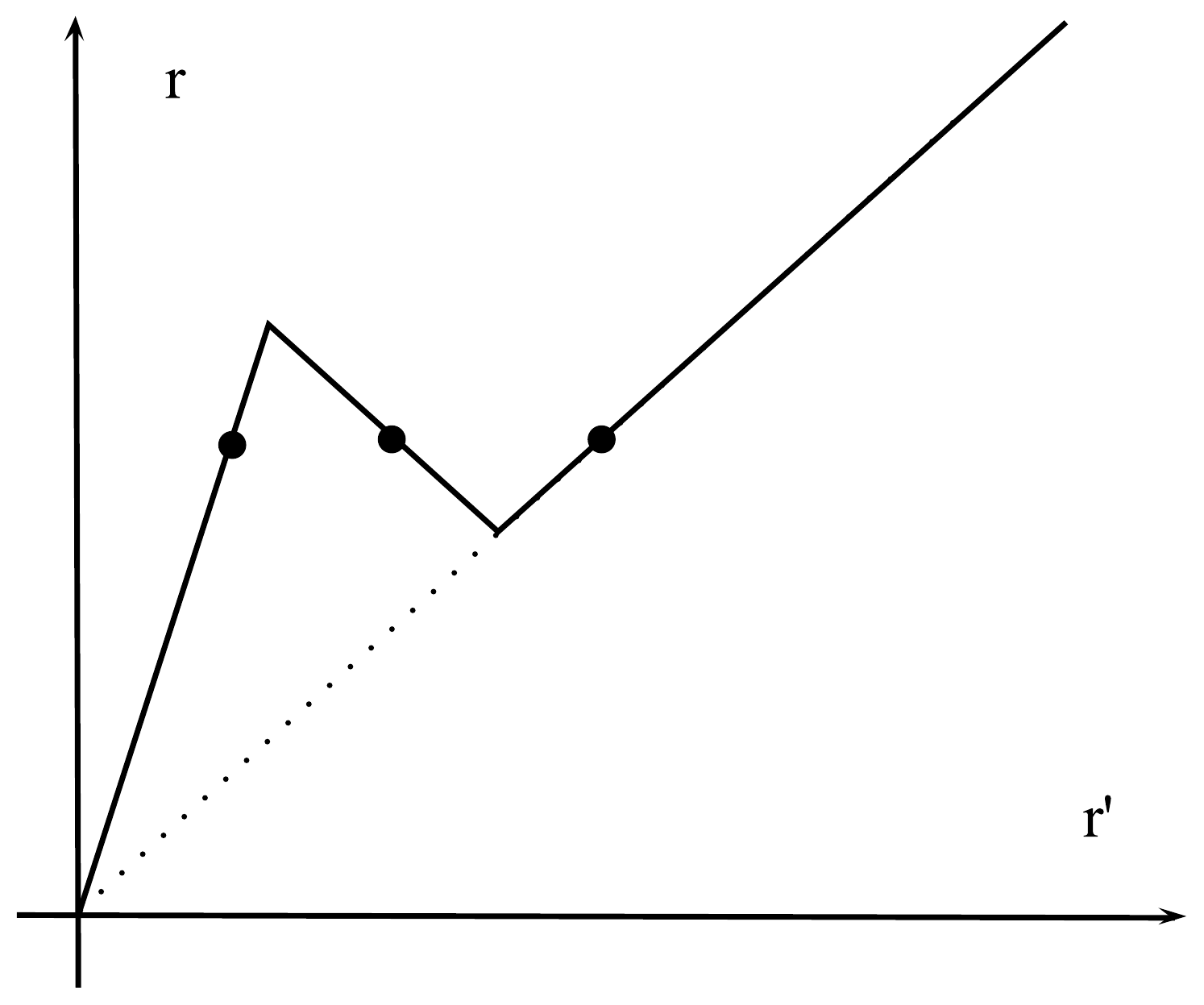}
\caption{\label{tfosup}   A superlens can be designed by folding the space on itself, here by transforming the radial distance. In this case, there is a part of the physical space that has a threefold image in the equivalent space. } }
\end{figure}

\begin{figure}[h]
{\centering \includegraphics[width=0.48\textwidth]{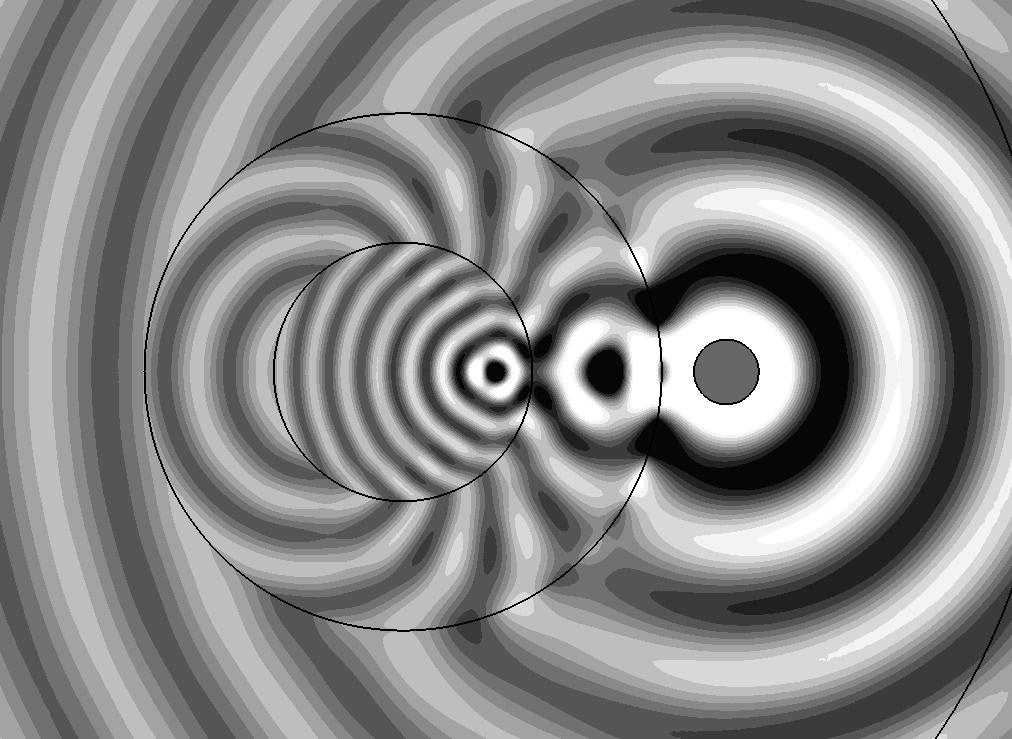}
\caption{\label{suplen}  A superlens designed by folding the empty space on itself do not perturbate the cylindrical waves emitted by a wire antenna but for the attenuation due to the dissipation introduced in the superlens permittivity in order to avoid the anomalous resonances \cite{anomalous}. The antenna has an image inside the superlens and inside the central part of the device.}}
\end{figure}

\begin{figure}[h]
{\centering  \includegraphics[width=0.48\textwidth]{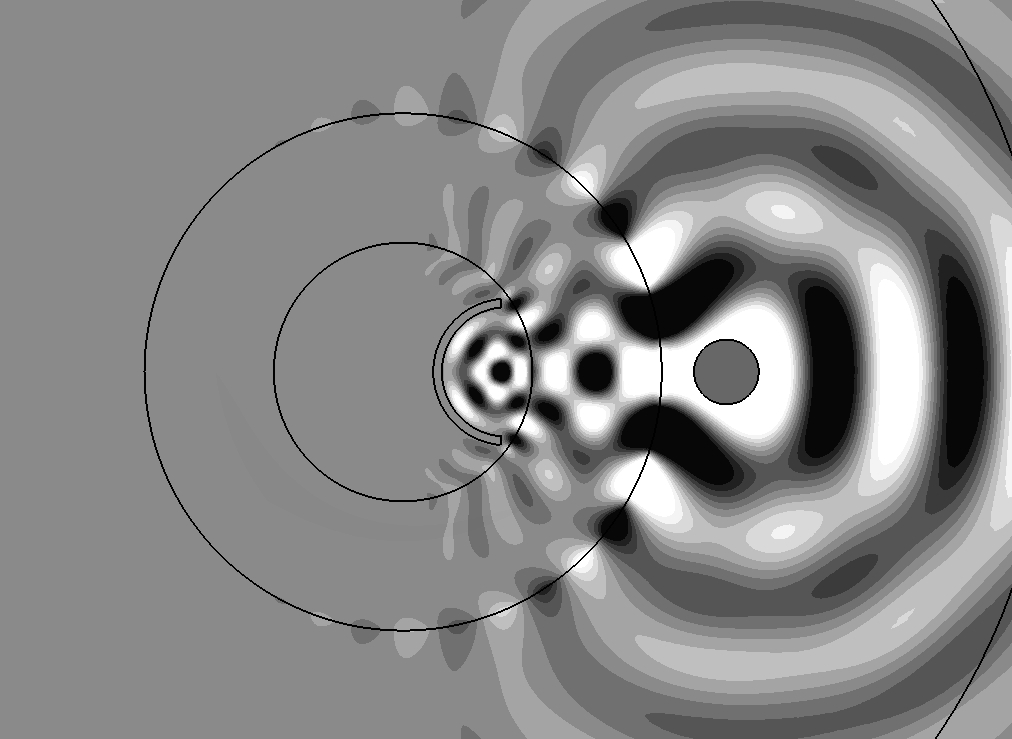}
\caption{\label{supdef} A small perfectly conducting deflector acts on the image of the antenna in the central part of the device and forces the waves to propagate only to the right. This can also be interpreted as if the deflector has a four time larger image acting on the original antenna.
 }}
\end{figure}

Another dramatic example of transformation optics devices are the superlenses \cite{suplens}: even if these devices
were proposed a few years before the rise of transformation optics, they are nicely interpreted as
corresponding to a folding of the space on itself. It has been suggested that such devices allow
a kind of ``remote action'' of the scatterers making possible things such as immaterial waveguides
called ``invisible tunnels'' \cite{invisi_tunnel}. They can also be used to set up a new kind of invisibility devices \cite{distantcloak}
and also illusion devices \cite{illusion} with a similar function that the one presented here with generalized cloaking but based on negative refraction index materials.  This device depends both on
the object to be transformed, since its scattering pattern has to be erased first by an \emph{ad hoc} ``antiobject'', and on the object to be faked while our device is more general since it depends only on the object to be faked independently of the
original object. As an illustration of superlensing, consider the multiple valued transformation of Fig. \ref{tfosup}. The part with negative slope corresponds to a negative refraction index material ($\underline{\underline{\varepsilon'}}$ and $\underline{\underline{\mu'}}$ have negative eigenvalues) and acts as a superlens. As a transformation of empty space, it does not perturbate the cylindrical waves emitted by a wire antenna (Fig. \ref{suplen} where the inner disk is the image of four time larger disk) but for the attenuation due to the dissipation introduced in the superlens permittivity in order to avoid the anomalous resonances \cite{anomalous}. One percent of losses has been added here to the value of the permittivity of the ideal perfect lens ($\underline{\underline{\varepsilon'}}$ has been multiplied by $1-0.01i$). The antenna on the right of the lens has two images, one inside the annular superlens and one inside the central part of the device so that we have well three copies of the antenna. In Fig. \ref{supdef}, a small perfectly conducting deflector inside the region surrounded by the perfect lens acts on the image of the antenna and forces the waves to propagate only to the right. This can also be interpreted as if the deflector has a four time larger image acting on the original antenna giving the illusion of a much larger object.

\section{Conclusion}

Transformation optics do not only offer the possibility to make optically disappear objects in invisibility cloaks but also to completely tune their optical signature i.e. to give them an arbitrary appearance.
   The possibility to place an object inside the coating of the cloak and that it will therefore appear different was already considered in \cite{opl} where a point source was shifted creating a mirage effect. A more general case is considered here since both the shape and the position of the object placed in the coating are modified.
 Note that if the object used to create the illusion is perfectly conducting and surrounds the central point of the cloak, its anamorphosis will surround the inner boundary of the cloak and it \emph{de facto} suppresses the singular behavior of the material properties. In this particular case,   the present approach is similar to the the hiding under the carpet idea of Li and Pendry \cite{carpet} but here a bounded object replaces the infinite reflecting plane. As an illustration of the state of the art cloaking technology, this latest device has been recently implemented in the realm of infrared using dielectric metamaterials \cite{dielectric_cloak}.
 Note also that our technique can be naturally extended to cloaks of arbitrary shapes \cite{elliptical,arbitrary}.


%

\appendix[Differential Geometry \cite{deschamps,burke,bossavit}\label{appendix}]

Given a $n$-dimensional space with a (global) co-ordinate system
$u_1, \cdots, u_n$ (not necessarily orthogonal), the exterior
derivative $d$ of a function $f(u_1, \cdots u_n)$ is its
differential $df=\sum_i \frac{\partial f}{\partial u_i}d u_i$. This
is a $1$-form. A general $1$-form can be written $\sum_i g_i(u_j) d
u_i $ where $g_i(u_j)$ are functions of the coordinates $u_j$. If a
$1$-form can be expressed as the differential of a function, it is
an exact $1$-form.

A curve $\gamma$ is an application from an interval $I=[t_0,t_1]$ of
$\R$ on
  the $n$-dimensional space: $\mathbf{r}(t)=(u_1(t), \cdots ,u_n(t))$ where
  $t$ is the parameter. The integral $\int_\gamma \alpha $ of a $1$-form $\alpha=\sum_i
g_i(u_j) d u_i$ on the curve $\gamma$ is defined by $\int_\gamma
\alpha = \int_{t_0}^{t_1} (\sum_i g_i(u_j(t)) \frac{\partial
u_i(t)}{\partial t})d t$. The value of the integral depends on $\gamma$ but does not depend
on the choice of the parameter.

The exterior product $\wedge $ is the skew-symmetric tensor product
such that $du_i \wedge du_j = -  du_j \wedge du_i = \frac{1}{2}(du_i
\otimes du_j  -  du_j \otimes du_i)$. A general $2$-form is a linear
combination $\sum_{i,j} g_{ij} du_i \wedge du_j $. The exterior
derivative of the $1$-form $\sum_i g_i d u_i $ is $d\sum_i g_i d u_i
= \sum_{i,j} \frac{\partial g_j}{\partial u_i} du_i \wedge du_j$.

A surface $\Sigma$ is an application from a two-dimensional open
domain $\Omega \subset \R^2$ on
  the $n$-dimensional space: $\mathbf{r}(s,t)=(u_1(s,t), \cdots ,u_n(s,t))$ where
  $s,t$ are the parameters. The integral $\int_\Sigma \beta $ of a $2$-form $\beta =$ $\sum_{i,j} g_{ij} du_i
\wedge du_j $ on the surface $\Sigma$ is defined by
$$\int_\Sigma \beta = \int\int_\Omega \sum_{i,j}
(g_{ij}\frac{\partial (u_i,u_j)}{\partial (s,t)}) ds dt$$ where
$\frac{\partial (u_i,u_j)}{\partial (s,t)}$ are the Jacobians. The
value of the surface (flux) integral depends on $\Sigma$ but does not depend on
the way the parameters are chosen.

The Stokes theorem states that $\int_\Sigma d\alpha = \int_{\partial
 \Sigma} \alpha$ where $\partial \Sigma$ is the boundary (curve) of
 the surface $\Sigma$.

More generally $p$-forms (with $0 \leq p \leq n$) are defined as
totally skew-symmetric tensors and can be manipulated using the exterior
derivative and the exterior product. Given a $1$-form $\alpha= \alpha_1 du_1
+\alpha_2 du_2 +\alpha_3 du_3 $ and a $2$-form $\beta= \beta_{23}
du_2 \wedge du_3 +\beta_{31} du_3 \wedge du_1+ \beta_{12} du_1
\wedge du_2  $, one has for instance: $d \alpha = (\frac{\partial\alpha_{2}}{\partial
u_1}-\frac{\partial\alpha_{2}}{\partial
u_1} )du_1 \wedge du_2 +  (\frac{\partial\alpha_{3}}{\partial
u_2}-\frac{\partial\alpha_{3}}{\partial
u_2} )du_2 \wedge du_3 +  (\frac{\partial\alpha_{1}}{\partial
u_3}-\frac{\partial\alpha_{1}}{\partial
u_3} )du_3 \wedge du_1 $, $d \beta= (
\frac{\partial\beta_{23}}{\partial
u_1}+\frac{\partial\beta_{31}}{\partial
u_2}+\frac{\partial\beta_{12}}{\partial u_3})du_1 \wedge du_2 \wedge
du_3$, and $\alpha \wedge \beta =
(\alpha_1\beta_{23}+\alpha_2\beta_{31}+\alpha_3\beta_{12} )du_1
\wedge du_2 \wedge du_3 $.

All the concepts here above rely only on the topological and differential structure
of the space.

The metric is a supplementary structure determined by a rank 2 covariant symmetric
tensor $\mathbf{g}$ whose $n^2$ coefficients form a positive
definite matrix. Given a metric, it is possible to introduce the concepts of scalar product, norm, distance, and angle.
The metric allows the
definition of a Hodge star operator $*$ that is a linear operator on
differential forms mapping $p$-forms on $(n-p)$-forms.

 Particular cases of spaces with a metric are the Euclidean
spaces $\mathbb{E}^n$ where Cartesian coordinates can be chosen so
that the coefficients of the metric form a unit matrix. For
$\mathbb{E}^3$, Cartesian coordinates are denoted
$\{u_1=x,u_2=y,u_3=z \}$ and the metric has the
  form:  $\mathbf{g}= dx \otimes dx + dy \otimes dy + dz \otimes dz
  $. In these Cartesian coordinates, the Hodge operator has the following
  action:\\
  $*dx= dy \wedge dz$, $*dy= dz \wedge dx$, $*dz= dx \wedge dy$\\
  $*(dx \wedge dy)= dz$, $*(dz \wedge dx)= dy$, $*(dy \wedge dz)=
  dx$\\ $*1 = dx \wedge dy \wedge dz $, $*(dx \wedge dy \wedge dz)= 1
  $.

\end{document}

%% file: mycommands.tex
\def\titreannexec4A{APPENDIX OF MULTIPOLE METHOD CHAPTER}

\newcommand{\Define}[2][]{%
\textit{#2}\ifthenelse{\equal{#1}{}}{\index{#2|textbf}}{\index{#1|textbf}}} 
\def\Capitalize#1#2|{{\uppercase {#1}}{#2}}

\def\v{\mathbf{v}}

\def\x{\mathbf{x}}

\def\build#1_#2^#3{\mathrel{\mathop{\kern 0pt#1}\limits_{#2}^{#3}}}

\def\rot{\mathop{\rm rot}\nolimits}

\def\R{\mathbb R}

\def\rot{\mathop{\rm \mathbf{curl}}\nolimits}

\def\p{\mathbf{p}}

\def\v{\mathbf{v}}

\def\x{\mathbf{x}}

\newcommand{\Cnp}[2]%
{%
{#2 \choose #1}
}

%% file: masking_compumag2009_jrnl_revised.bbl
\begin{thebibliography}{1}

\bibitem{pendrycloak} J.B. Pendry, D. Shurig, D.R. Smith,
"Controlling electromagnetic fields",
\newblock Science {\bf 312}, 1780 (2006).

\bibitem{jkrowling} J.K. Rowling, \emph{Harry Potter and the Chamber of Secrets}, (Bloomsbury Publishing PLC, 1998).

\bibitem{opl} F. Zolla, S. Guenneau, A. Nicolet, J. B. Pendry,
"Electromagnetic analysis of cylindrical invisibility cloaks and the mirage effect",
\newblock  Opt. Lett. \textbf{32}, 1069 (2007).

\bibitem{nicolet} A. Nicolet, J. F. Remacle, B. Meys, A. Genon, W. Legros,
"Transformation methods in computational electromagnetism",
 J. Appl. Phys. {\bf 75}, 6036 (1994).

\bibitem{deschamps} G. A. Deschamps,"Electromagnetics and differential forms", \newblock Proc. IEEE,
\textbf{69}, 676 (1981).

\bibitem{burke} W. L. Burke, \emph{Applied Differential Geometry}, (Cambridge University Press, 1985).

\bibitem{symplectic} V. Guillemin, S. Sternberg, \emph{Symplectic Techniques in Physics}, (Cambridge University Press, 1990).

\bibitem{henrotte} F. Henrotte, B. Meys, H. Hedia, P. Dular, and W. Legros, "Finite element modelling with transformation techniques",
IEEE Trans. Mag. \textbf{35}, 1434, (1999).

\bibitem{bossavit} A. Bossavit, J. Japan Soc. Appl. Electromagn.
\& Mech., \textbf{6}, "On the geometry of electromagnetism. (1): Euclidean space", 17, "On the geometry of electromagnetism. (2): Geometrical objects", 114, "On the geometry of electromagnetism. (3): Faraday's law", 233, "On the geometry of electromagnetism. (4): 'Maxwell's house' ", 318 (1998), \textbf{7}, "Computational electromagnetism and geometry: Building a finitedimensional
'Maxwell's house'. (1): Network equations", 150, "Computational electromagnetism and geometry. (2): Network
constitutive laws", 294, "Computational electromagnetism and geometry. (3): Convergence", 401 (1999), \textbf{8}, "Computational electromagnetism and geometry. (4): From degrees of
freedom to fields", 102, "Computational electromagnetism and geometry. (5): The 'Galerkin
hodge'", 203, "Computational electromagnetism and geometry. (6): Some questions
and answers", 372 (2000).

\bibitem{pcfbook} F. Zolla, G. Renversez, A. Nicolet, B. Kuhlmey, S.
Guenneau, D. Felbacq, \newblock \emph{Foundations of Photonic Crystal Fibres},
(Imperial College Press, 2005).

\bibitem{twistedEPJ}   A. Nicolet, F. Zolla, S. Guenneau, "Modelling of twisted optical waveguides with edge elements", Eur. Phys. J. - Appl. Phys. {\bf 28}, 153 (2004).

\bibitem{twistedJWAves} A. Nicolet, F. Zolla, Y. Ould Agha, S. Guenneau, "Leaky modes in twisted microstructured optical fibers",  Waves in Random and Complex Media {\bf 17}, 559 (2007).

\bibitem{milton} G. W. Milton, M. Briane, J. R. Willis,
"On cloaking for elasticity and physical equations with a transformation invariant form",
\newblock New J. of Phys.
\textbf{8}, 248 (2006).

\bibitem{compel_geo}
A. Nicolet, F. Zolla, Y. Ould Agha, S. Guenneau, "Geometrical transformations and equivalent materials in computational electromagnetism",
\newblock COMPEL \textbf{27}, 806 (2008).

\bibitem{teixeira}
F. L. Teixeira, "Differential form approach to
the analysis of electromagnetic
cloaking and masking", Microwave Opt. Technol. Lett. {\bf 49},  2051 (2007).

\bibitem{getdp}
P. Dular, C. Geuzaine, F. Henrotte, W. Legros,
\newblock "A general environment for the treatment of discrete problems and its application to the finite element method",
 IEEE Trans. Mag.  { \bf 34},
3395 (1998). (see also http://www.geuz.org/getdp/)

\bibitem{gmsh}
C. Geuzaine, J.-F. Remacle,
"Gmsh: a three-dimensional finite element mesh generator with built-in pre- and post-processing facilities",
Int. J. Numer. Methods Eng.
 \textbf{79},
1309 (2009).

\bibitem{suplens} J.B. Pendry, J. B.,  "Negative refraction makes a perfect lens", Phys. Rev. Lett.  {\bf 85}, 3966 (2000).

\bibitem{invisi_tunnel}  Zhang, J.J., Y. Luo, H. S. Chen, J. Huangfu, B.-I. Wu, L. Ran, J. A. Kong,
"Guiding waves through an invisible tunnel",
\newblock  Opt. Exp. \textbf{17}, 6203 (2009).

\bibitem{distantcloak} Yun Lai, Huanyang Chen, Zhao-Qing Zhang, and C. T. Chan, "Complementary Media Invisibility Cloak that Cloaks Objects at a Distance Outside the Cloaking Shell", Phys. Rev. Lett. {\bf 102}, 093901 (2009).

\bibitem{illusion}  Yun Lai, Jack Ng, HuanYang Chen, DeZhuan Han, JunJun Xiao, Zhao-Qing Zhang, and C. T. Chan,
"Illusion Optics: The Optical Transformation of an Object into Another Object", Phys. Rev. Lett. {\bf 102}, 253902 (2009).

\bibitem{anomalous}  G.W. Milton,  N.-A.P. Nicorovici,  "On the cloaking effects associated with anomalous localized resonance" Proc. R. Soc. Lond. A {\bf 462} 3027 (2006)

\bibitem{carpet}
 J.S. Li, J.B. Pendry,
"Hiding under the Carpet: A New Strategy for Cloaking",
\newblock  Phys. Rev. Lett. \textbf{101}, 203901 (2008)

\bibitem{dielectric_cloak}
J. Valentine, J. Li, T. Zentgraf, G. Bartal, X. Zhang, "An optical cloak made of dielectrics", Nature Materials, \textbf{8}, 568 (2009).

\bibitem{elliptical}
A. Nicolet, F. Zolla, S. Guenneau, "Finite element analysis of cylindrical invisibility cloaks of elliptical cross section", IEEE Trans. Mag. \textbf{44}, 1153 (2008).

\bibitem{arbitrary}
 A. Nicolet, F. Zolla, S. Guenneau, "Electromagnetic analysis of cylindrical cloaks of an arbitrary cross section",
\newblock  Opt. Lett. \textbf{33}, 1584 (2008).



\end{thebibliography}
